\documentstyle[preprint,aps,epsf]{revtex}
\begin{document}
\tightenlines
\draft
\newcommand{\be}{\begin{equation}}
\newcommand{\ee}{\end{equation}}
\newcommand{\bea}{\begin{eqnarray}}
\newcommand{\eea}{\end{eqnarray}}

\title{Exact results for the zeros of the partition function of 
the Potts model on finite lattices}
\author{Seung-Yeon Kim\footnote{E-mail: kim@cosmos.psc.sc.edu}
and Richard J. Creswick\footnote{Corresponding author.
E-mail: creswick.rj@sc.edu}}
\address{Department of Physics and Astronomy,
University of South Carolina,\\
Columbia, South Carolina 29208, USA}
\maketitle

\begin{abstract}
The Yang-Lee zeros of the $Q$-state Potts model are investigated
in 1, 2 and 3 dimensions. Analytical results derived from the transfer
matrix for the one-dimensional model reveal a systematic behavior
of the locus of zeros as a function of $Q$. For $1<Q<2$ the zeros 
in the complex $x=\exp(\beta H_q)$ plane lie inside the unit circle,
while for $Q>2$ they lie outside the unit circle for finite temperature. 
In the special case $Q=2$ the zeros lie exactly
on the unit circle as proved by Lee and Yang. 
In two and three dimensions the zeros are calculated numerically 
and behave in the same way.
Results are also presented for the critical line of the Potts model
in an external field as determined from the zeros of the partition
function in the complex temperature plane.
\end{abstract}

\section{introduction}

The $Q$-state Potts model\cite{potts} in two and three dimensions
exhibits a rich variety of critical behavior
and is very fertile ground for the analytical and numerical
investigation of first- and second-order phase transitions.
With the exception of the two-dimensional $Q=2$ Potts (Ising) model
in the absence of an external magnetic field\cite{onsager}, 
exact solutions for arbitrary $Q$ are not known.
However, some exact results have been established for the
two-dimensional $Q$-state Potts model.
For $Q\le4$ there is a second-order phase transition, while for $Q>4$
the transition is first order\cite{baxter1}.
From the duality relation the critical temperature
is known to be $T_c=J/k_B\ln(1+\sqrt{Q})$\cite{potts}.
For $Q\le4$ the critical exponents\cite{creswick1}
are known, while for $Q > 4$ the latent heat\cite{baxter1}, 
spontaneous magnetization\cite{baxter2},
and correlation length\cite{buffernoir} at $T_c$ are also known.

The $Q$-state Potts model on a lattice $G$ 
in an external magnetic field $H_q$
is defined by the Hamiltonian
\be
{\cal H}_Q=J\sum_{\langle i,j\rangle}[1-\delta(\sigma_i,\sigma_j)]-H_q\sum_k\delta(\sigma_k,q),
\ee
where $J$ is the coupling constant, 
$\langle i,j\rangle$ indicates a sum over nearest-neighbor pairs,
$\sigma_i=1,...,Q$,
and $q$ is a fixed integer between 1 and $Q$.
The partition function of the model can be written as 
\be 
Z_Q(x,y)=\sum_{M=0}^{N_s}\sum_{E=0}^{N_b}\Omega_Q(M,E) x^M y^E,
\ee
where $x=e^{\beta H_q}$, $y=a^{-1}=e^{-\beta J}$,
$E$ and $M$ are positive integers $0\le E\le N_b$ and $0\le M\le N_s$,
respectively, $N_b$ and $N_s$ are the number of bonds
and the number of sites on the lattice $G$,
and $\Omega_Q(M,E)$ is the number of states
with fixed $E$ and fixed $M$.
From Eq. (2) it is clear that $Z_Q(x,y)$ is simply a polynomial 
in $x$ and $y$.   

By introducing the concept of the zeros of the partition function
in the {\it complex} magnetic-field plane (Yang-Lee zeros),
Yang and Lee\cite{yang} proposed a mechanism
for the occurrence of phase transitions in the thermodynamic limit
and yielded a new insight into the unsolved problem of the two-dimensional 
Ising model in an arbitrary nonzero external magnetic field. 
It has been shown\cite{yang,creswick2} that the distribution of the zeros 
of a model determines its critical behavior.
Lee and Yang\cite{yang} also 
formulated the celebrated circle theorem which states that
the Yang-Lee zeros of the Ising ferromagnet lie on the unit circle.
While we lack the circle theorem of Lee and Yang to tell us the location
of the zeros for $Q\ne2$, something can be said about their general behavior
as a function of temperature.
At zero temperature ($y=0$) from Eq. (2) the partition function is
\bea
Z_Q(x,0)&=&\sum_{M}\Omega_Q(M,0)x^M \cr
&=&(Q-1)+x^{N_s}.
\eea
Therefore, the Yang-Lee zeros at $T=0$ are given by
\be
x_k=(Q-1)^{1/N_s}\exp[i(2k-1)\pi /N_s],
\ee
where $k=1,...,N_s$.
The zeros at $T=0$ are uniformly distributed on the
circle with radius $(Q-1)^{1/N_s}$
which approaches unity in the thermodynamic limit, independent of $Q$.
At infinite temperature ($y=1$) Eq. (2) becomes
\be
Z_Q(x,1)=\sum_{M=0}^{N_s}\sum_{E=0}^{N_b}\Omega_Q(M,E)x^M.$$
\ee
Because $\sum_{E}\Omega_Q(M,E)$ is simply
${N_s\choose M}(Q-1)^{N_s-M}$,
at $T=\infty$, the partition function is given by
\be
Z_Q(x,1)=(Q-1+x)^{N_s},
\ee
and its zeros are $N_s$-degenerate at $x=1-Q$,
independent of lattice size.

\section{Yang-Lee zeros in one dimension}

For the one-dimensional Potts model in an external field the eigenvalues 
of the transfer matrix were found by Glumac and Uzelac\cite{glumac}.
The two dominant eigenvalues are $\lambda_\pm=(A\pm i B)/2a$,
where $A=a(1+x)+Q-2$, $B=-i\sqrt{[a(1-x)+Q-2]^2+4(Q-1)x}$,
and $\lambda_0=(a-1)/a$ is $(Q-2)$-fold degenerate. 

We will assume that $|\lambda_\pm| > \lambda_0$ and verify this
assumption a posteriori. The partition function is
\be
Z_N=\lambda_+^N+\lambda_-^N+(Q-2)\lambda_0^N,
\ee
but, by the above approximation, for large $N$ we have
\be
Z_N\simeq\lambda_+^N+\lambda_-^N.
\ee 
If we define $A=2C\cos\psi$ and $B=2C\sin\psi$, 
where $C=\sqrt{(a-1)(Q+a-1)x}$, then $\lambda_\pm=(C/a)\exp(\pm i\psi)$,
and the partition function is
\be
Z_N=2\bigg({C\over a}\bigg)^N \cos N\psi.
\ee
The zeros of the partition function are then given by
\be
\psi=\psi_k={2k+1\over2N}\pi,\ \ \ \ \ k=0,1,2,...,N-1.
\ee
The location of these zeros in the complex $x$ plane is determined
by the solution of 
\be
A=2C\cos\psi_k.
\ee
This equation is quadratic in $z=\sqrt{x}$ with roots
\be
z_k={1\over a}\bigg[\sqrt{(a-1)(Q+a-1)}\cos\psi_k\pm i
\sqrt{(a-1)(Q+a-1)\sin^2\psi_k+Q-1}\bigg].
\ee
It is easily verified that
\be
|z_k|^2=|x_k|={Q+a-2\over a},
\ee
and we see that all the zeros lie on a circle in the complex $x$ plane.
The argument of $x_k$ is given by
\be
\cos{\theta_k\over2}=\sqrt{(a-1)(Q+a-1)\over a(Q+a-2)}\cos\psi_k.
\ee

Before we analyze these results we must verify that in fact
$|\lambda_\pm|>\lambda_0$ in the region of these zeros.
We find
\be
{|\lambda_\pm|\over\lambda_0}=\sqrt{(Q+a-1)(Q+a-2)\over a(a-1)}
\ee
which is indeed greater than unity for $Q>1$.

If we return to Eq. (13) we can discern a remarkably systematic behavior
of the Yang-Lee zeros with $Q$. For $1<Q<2$, $|x_k|<1$ and the zeros
lie inside the unit circle. As $a\to\infty$, the zeros approach the unit 
circle from within, as we observed in Eq. (4), and for $a=1$,
$\cos(\theta_k/2)=0$, so $\theta_k=\pi$ and all the zeros lie at $1-Q$,
as predicted in Eq. (6).
For $Q>2$, $|x_k|>1$ and the zeros lie outside the unit circle
and approach the unit circle as $a\to\infty$ from outside.
In the special case $Q=2$ we of course find that $|x_k|=1$,
as proved by Lee and Yang\cite{yang}.
The edge singularity in the thermodynamic limit is given by 
\be
\theta_0=2\cos^{-1}\sqrt{(a-1)(Q+a-1)\over a(Q+a-2)}>0,
\ee
from which we conclude that no transition occurs for any $T>0$.

Finally, we can use these exact results to study finite size effects
on the distribution of zeros. For finite $N$ we must find the zeros of 
the full partition function as given in Eq. (7).
We find that finite size effects are very small in one dimension,
but the largest deviation from the limiting locus occurs for
$\arg(x)=\pi$.

\section{Yang-Lee zeros in two and three dimensions}

We have calculated {\it exact} integer values for 
$\Omega_Q(M,E)$ on $L\times L$ square lattice for $3\le L\le8$ and
$3^3$ simple-cubic lattice ($Q=3$)
using the restricted microcanonical transfer matrix\cite{creswick3}.
For square lattices up to $L=12$ ($Q=3$) the zeros are calculated 
using a {\it semi-canonical} variation of the transfer matrix.
Fig. 1a shows the Yang-Lee zeros
for the two-dimensional three-state Potts model in the complex $x$ plane
at the critical temperature $y_c=1/(1+\sqrt{3})=0.366...$
for $L=4$ and $L=10$ with cylindrical boundary conditions.
Note that just as in the one-dimensional model, the zeros of the three-state
Potts model lie close to, but not on, the unit circle.
The zero farthest from the unit circle is in the neighborhood
of $\arg(x)=\pi$, while the zero closest to the positive 
real axis lies closest to the unit circle.
The behavior of the zeros with the size of the lattice also follows
that of the one-dimensional model;
the zeros for $L=10$ lie on a locus interior to
that for $L=4$, and we find similar behavior for larger values of $Q$.
In the thermodynamic limit the locus of zeros cuts
the real $x$ axis at the point $x=1$\cite{kim1} corresponding to $H_q=0$,
as described by Yang and Lee\cite{yang}. 
Fig. 1b shows the zeros for the two-dimensional three-state Potts model 
at several temperatures with cylindrical boundary conditions.
At $y=0.5y_c$ the zeros are nearly uniformly distributed 
in argument and close to the unit circle. 
As the temperature is increased the edge singularity 
moves away from the real axis and the zeros detach  from the unit circle.
Finally, as $y$ approaches unity, the zeros converge on the point $x=-2$.
As predicted by Eqs. (4) and (6),  
for periodic boundary conditions\cite{kim2} and self-dual boundary 
conditions\cite{kim1} we observe the same behaviors as those
in Fig. 1 for cylindrical boundary conditions.

The exact nature of the locus of zeros for the $Q$-state Potts model
in two dimensions is unknown, with the exception of $Q=2$.
It is clear that the Yang-Lee zeros of the two-dimensional $Q$-state
Potts model do not lie on the unit circle for $Q>2$ 
for any value of $y$ and any finite value of $L$. 
Since the zero in the neighborhood of $\arg(x)=\pi$ is always
the farthest from the unit circle, if this zero can be shown to
approach $|x(\pi)|=1$ in the limit $L\to\infty$, all the zeros 
should lie on the unit circle in this limit. In Fig. 2 we show
values for $|x(\pi)|$ extrapolated to infinite size using the Bulirsch-Stoer 
(BST) algorithm\cite{bst} for $3\le Q\le8$ at $y=0.5y_c$ and $y=y_c$.
The error bars are twice the difference 
between the $(n-1,1)$ and $(n-1,2)$ approximants.
From these results it is clear that while the locus of zeros lies 
{\it close} to the unit circle at $y=y_c$, it does not coincide
with it except at the critical point $x=1$. 
While the numerical evidence presented here suggests that the locus 
of zeros for the three-state model is not the unit circle,
and the exact result in one-dimension provides an example where
the locus varies continuously with $Q$, the nature of the locus remains
an open and fascinating question. 

Fig. 3 shows the BST estimate of the 
modulus of the locus of zeros as a function of angle for 
the two-dimensional three-state Potts model 
at $y=0.5y_c$, $y=y_c$ and $y=1.2y_c$.
To calculate the extrapolated values for each angle, $\theta$,
we selected the zero whose arguments were closest to $\theta$
for lattices of size $3 \le L \le 12$
for $\theta=0.0, 0.5,...,2.5$ and $\pi$. 
The BST algorithm was then used to extrapolate 
these values for finite lattices to infinite size.
The large variation in the size of the error bars is due to
the fact that for a given $\theta$ there may be no zero 
{\it close} to $\theta$ for the smaller lattices.  
In Fig. 3 at $y=y_c$ the first four angles are shifted slightly
from the original values ($\theta=0.0$, 0.5, 1.0 and 1.5)
to be distinguished from the results at $y=0.5y_c$.
For $y=0.5y_c$ and $y=y_c$ the first zeros definitely lie on the
point $r(\theta=0)=1$ in the thermodynamic limit.
However, for $y=1.2y_c$ the BST estimates of the modulus and angle of the
first zero are 1.054(2) and 0.09(6).
Therefore, at $y=1.2y_c$ the locus of zeros does not cut the
positive real axis in the thermodynamic limit,
consistent with the absence of a physical singularity for $y > y_c$.

Fig. 4 shows the Yang-Lee zeros for the three-dimensional three-state Potts
model at several temperatures on the $3^3$ simple-cubic lattices
with periodic boundary conditions in $x$ and $y$ directions
and free boundary conditions in $z$ direction.
The behaviors in Fig. 4 for the three-dimensional model are the same as 
those in Fig. 1b for the two-dimensional model.

\section{Fisher zeros in an external field}

Fisher\cite{fisher} emphasized that the partition
function zeros in the complex temperature plane (Fisher zeros)
are also very useful in understanding phase transitions.
In particular, using the Fisher zeros both the ferromagnetic
phase and the antiferromagnetic phase can be considered
at the same time.
From the exact solutions\cite{onsager} of the square lattice Ising model
it has been shown\cite{fisher} that in the absence of an external magnetic 
field the Fisher zeros lie on two circles in the thermodynamic limit.
Recently the locus of the Fisher zeros of the $Q$-state Potts model
in the absence of an external magnetic field
has been studied extensively\cite{bhanot,chen}.
It has been shown\cite{chen}
that for self-dual boundary conditions 
near the ferromagnetic critical point $y_c=1/(1+\sqrt{Q})$
the Fisher zeros of the Potts model on a finite square lattice
lie on the circle with center $-1/(Q-1)$ and radius $\sqrt{Q}/(Q-1)$
in the complex $y$ plane, while the antiferromagnetic circle 
of the Ising model completely disappears for $Q>2$.  
However, the properties of the Fisher zeros for $Q>2$
in an external field are not known.

In the limit $H_q\to-\infty$ ($x\to0$) the partition function 
of the $Q$-state Potts model becomes
\be 
Z_Q=\sum_{E=0}^{N_b}\Omega_Q(M=0,E) y^E,
\ee
where $\Omega_Q(M=0,E)$ is the same as the number of states 
$\Omega_{Q-1}(E)$ of the ($Q-1$)-state Potts model 
in the absence of an external magnetic field.
As $x$ decreases from 1 to 0, the $Q$-state Potts model is transformed
into the ($Q-1$)-state Potts model in zero external field\cite{kim3}. 
For an external field $H_q<0$, one of the Potts states is supressed
relative to the others. The symmetry of the Hamiltonian is that of the
$(Q-1)$-state Potts model in zero external field, so that we expect
to see cross-over from the $Q$-state critical point to the $(Q-1)$-state
critical point as $-H_q$ is increased.

We have studied the field dependence of the critical point for 
$0\le x\le1$ through the Fisher zero closest to the real axis, $y_1(x,L)$,
for the two-dimensional three-state Potts model.
For a given applied field $y_1$ approaches the critical point
$y_1(L)\to y_c(x)$ in the limit $L\to\infty$, and the thermal 
exponent $y_t(L)$ defined as\cite{creswick3,bhanot}
\be
y_t(L)=-{\ln \{{\rm Im}[y_1(L+1)]/{\rm Im}[y_1(L)]\}\over\ln[(L+1)/L]}
\ee
will approach the critical exponent $y_t(x)$.
Table I shows values for $y_c(x)$ extrapolated from calculations of 
$y_1(x,L)$ on $L\times L$ lattices for $3\le L\le8$ using the BST algorithm.
The critical points for $x=1$ (three-state) and $x=0$ (two-state) 
Potts models are known exactly and are included in Table I for comparison.
Note that the imaginary parts of $y_c$(BST) are all consistent with zero.
We have also calculated the thermal exponent, $y_t$, applying
the BST algorithm to the values given by Eq. (18), and these results 
are also presented in Table I. For $x=1$ we find $y_t$
very close to the known value $y_t=6/5$ for the three-state model,
but for $x$ as large as 0.5 we obtain $y_t=1$, the value of the 
thermal exponent for the two-state (Ising) model. 

Fig. 5 shows the critical line of the two-dimensional three-state Potts 
ferromagnet for $H_q < 0$.
In Fig. 5 the upper line is the critical temperature of the two-state model, 
$T_c(Q=2)=1/\ln(1+\sqrt{2})$, 
and the lower line is the critical temperature for the three-state model,
$T_c(Q=3)=1/\ln(1+\sqrt{3})$.
The critical line for small $-H_q$ is given by
$T-T_c(Q=3)\sim(-H_q)^{y_t/y_h},$
where $y_t=6/5$ and $y_h=28/15$ for the three-state Potts model.

\section{conclusion}

Following the work of Glumac and Uzelac\cite{glumac} we have studied
the locus of Yang-Lee zeros in the complex $x$ plane for the $Q$-state Potts
model in one-dimension and find that for $Q>1$ the zeros lie on a circle 
whose radius varies continuously with temperature and $Q$. For $1<Q<2$ 
and any finite temperature, the zeros lie inside the unit circle,   
while for $Q>2$ they lie outside the unit circle. In the special case
$Q=2$ (Ising model) the zeros lie exactly on the unit circle, as first
proved by Lee and Yang\cite{yang}.

In two and three dimensions we have used the microcanonical transfer
matrix\cite{creswick3} to find the Yang-Lee zeros for finite lattices. 
The general trends observed in one dimension are repeated in two and
three dimensions. In two dimensions with $3\le Q\le8$ the zeros 
lie outside the unit circle, and finite-size scaling suggests that 
the locus of zeros in the thermodynamic limit touches the unit circle
only at the critical point, $x=1$. In three dimensions with $Q=3$ we 
have calculated the zeros on a single $3\times3\times3$ lattice,
and there again the zeros lie outside the unit circle. While the exact 
form of the locus of zeros remains an open and important question,
our results suggest a universal behavior which incorporates the Lee-Yang
circle theorem. For $1<Q<2$ the locus of zeros lies within the unit circle
while for $Q>2$ it lies outside the unit circle and in each case
touches the unit circle at the critical point $x=1$.

We have also studied the Fisher (complex temperature) zeros 
of the two-dimensional three-state Potts model for lattices of 
size $3\le L\le8$ in the presence of an external field. 
The field reduces the symmetry
of the Hamiltonian to that of the ($Q-1$)-state model, and by studying
the edge singularity we are able to determine the critical point
and temperature exponent as a function of the field.
Cross-over from the 3-state to the 2-state universality class is 
apparent. A more detailed analysis using larger lattices and
allowing for confluent singularities is in progress.

\begin{table}
\caption{The critical temperature $y_c$ and the critical exponent $y_t$ 
of the two-dimensional three-state Potts model for $0\le x\le1$.}
\begin{tabular}{ccclc}
$x$ &$y_c$ (BST) &$y_c$ (exact) &$y_t$ (BST) &$y_t$ (exact) \\
\hline
0     &$0.414(3)+0.0002(4)i$ &0.414213... &1.001(2) &1   \\
0.001 &$0.414(3)+0.0002(4)i$ &            &1.001(2)      \\
0.05  &$0.413(5)+0.0002(3)i$ &            &1.0009(6)     \\
0.5   &$0.400(2)+0.000(2)i$  &            &0.982(21)     \\
1     &$0.366(2)+0.0002(5)i$ &0.366025... &1.195(3) &$6\over5$ \\
\end{tabular}
\end{table}

\begin{figure}
\epsfbox{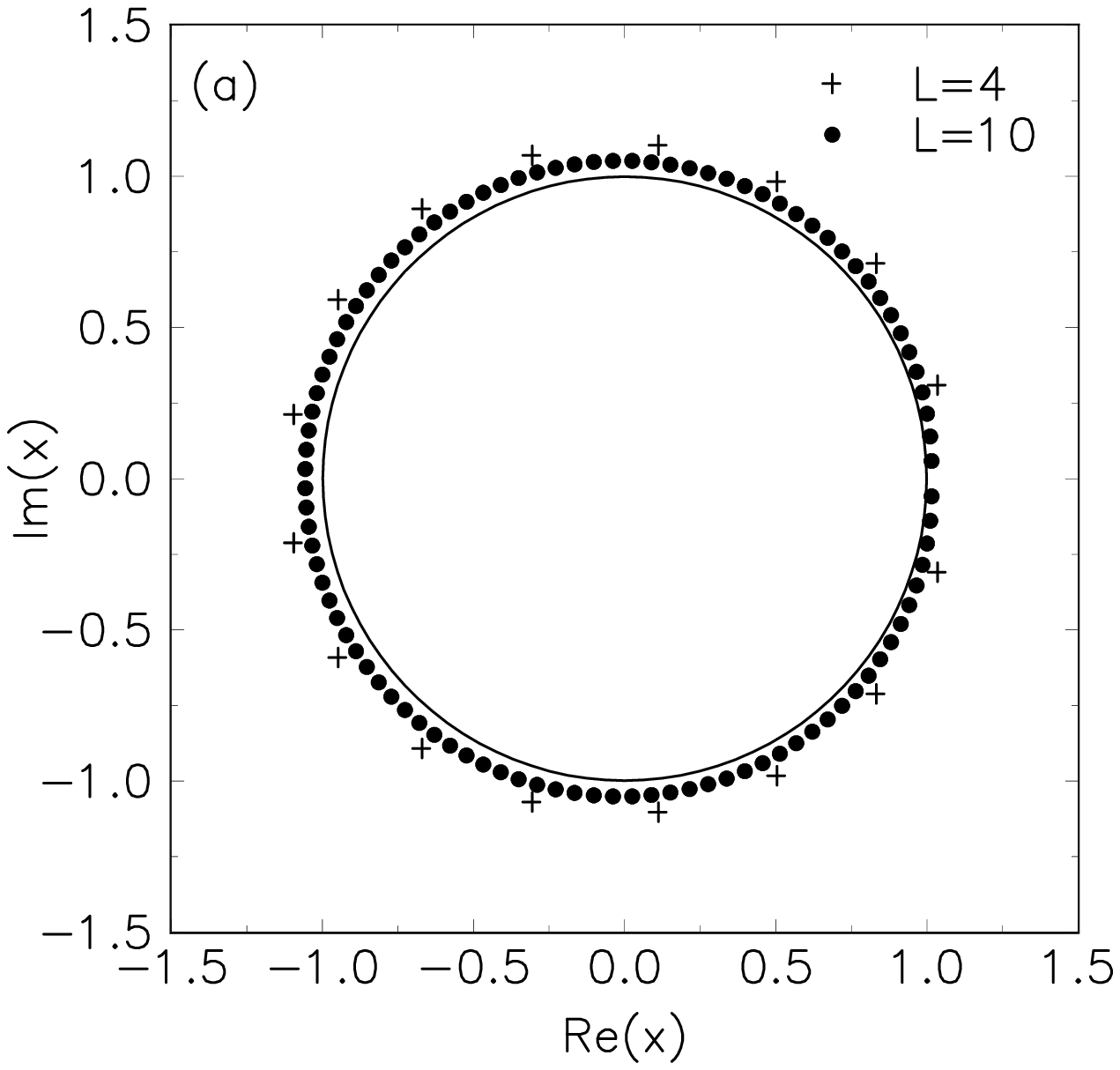}
\epsfbox{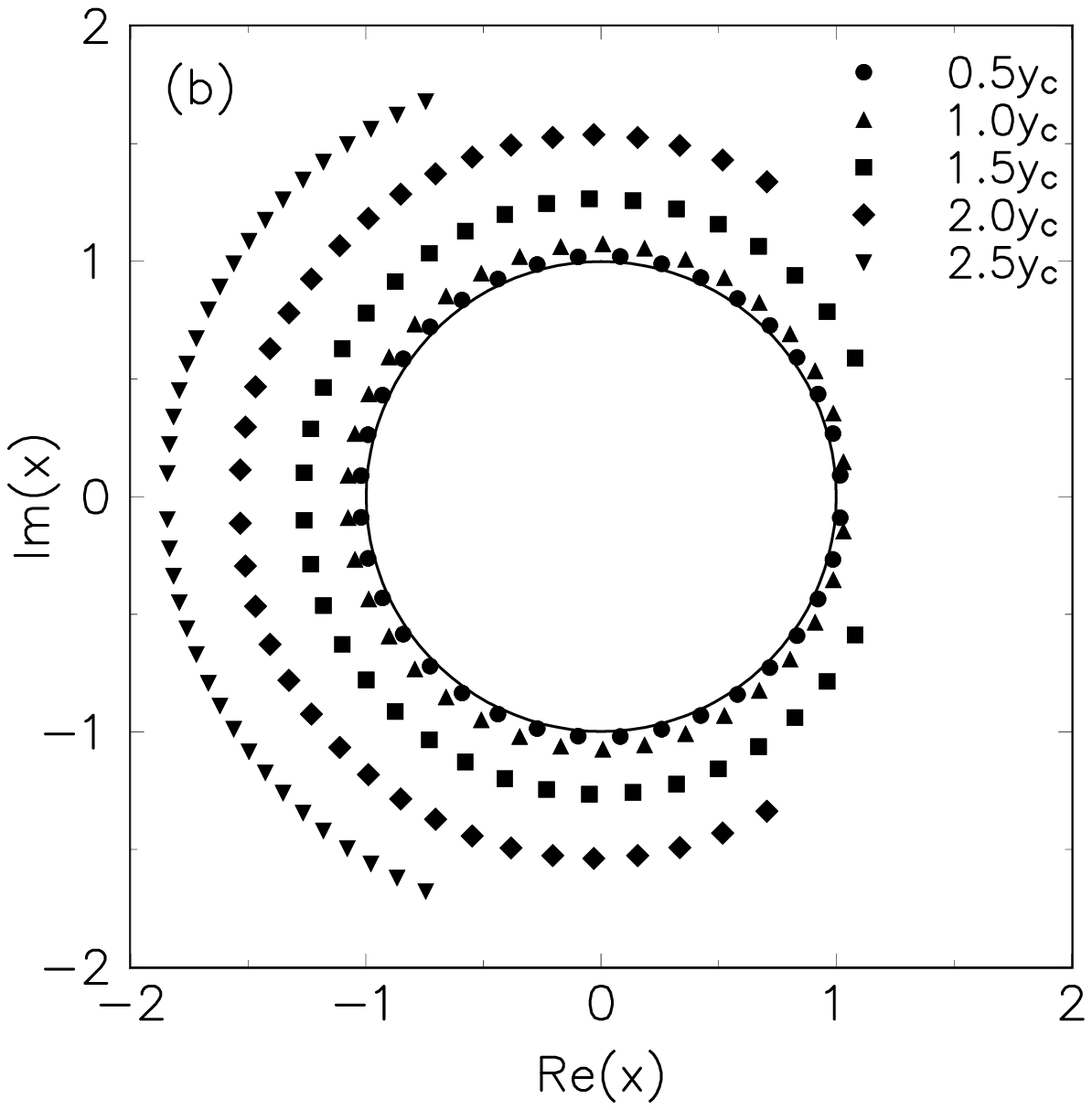}
\caption{Zeros of the two-dimensional three-state Potts model in the 
complex $x$ plane with cylindrical boundary conditions
(a) at $y=y_c$ for $L=4$ and $L=10$ 
and (b) for several values of $y$ ($L=6$)}
\end{figure}

\begin{figure}
\epsfbox{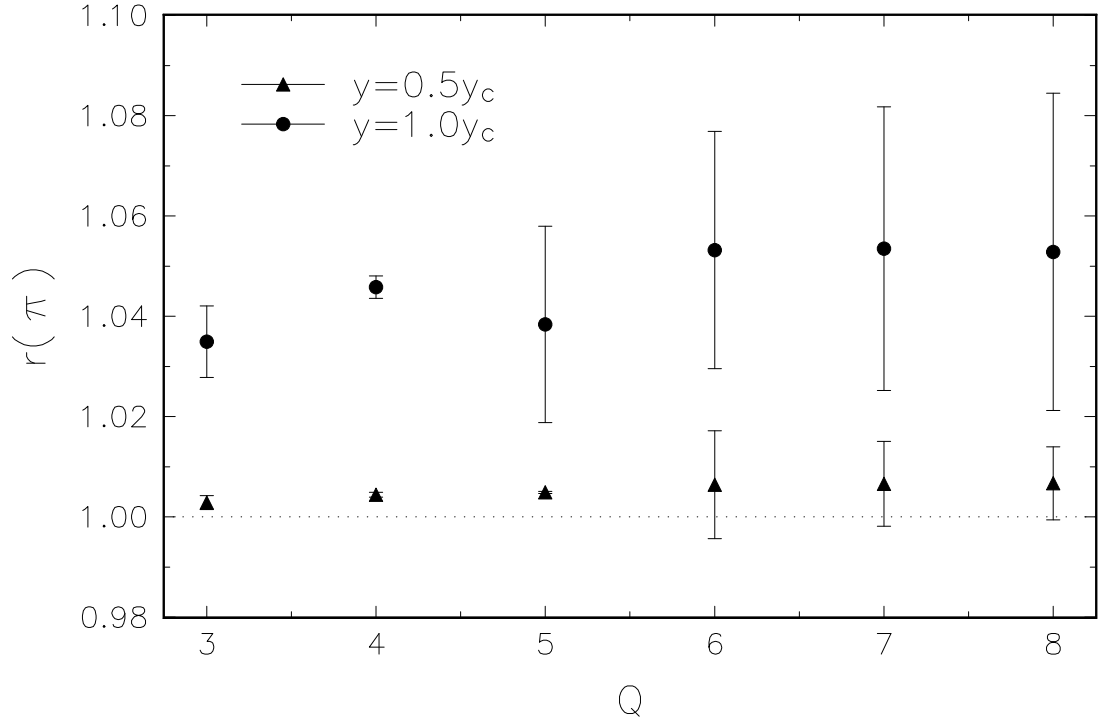}
\caption{Modulus of the zero at $\theta=\pi$
extrapolated to infinite size for $3\le Q\le8$ at 
$y=0.5y_c$ and $y=y_c$ with cylindrical boundary conditions.}
\end{figure}

\begin{figure}
\epsfbox{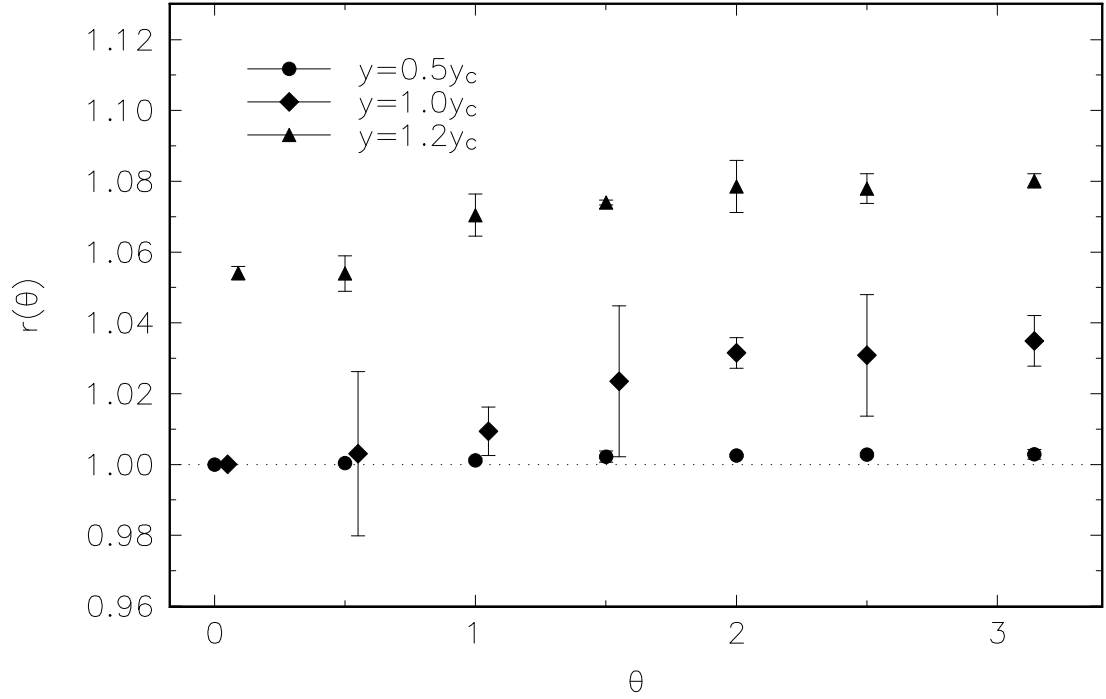}
\caption{Modulus of the locus of zeros as a function of angle for  
the two-dimensional three-state Potts model at $y=0.5y_c$, $y_c$, 
and $1.2y_c$ with cylindrical boundary conditions.
The slight horizontal off-set for data for $y=y_c$ is for clarity only.
However, the off-set of the edge singularity for $y=1.2y_c$ from
$\theta=0$ is real.}
\end{figure}

\begin{figure}
\epsfbox{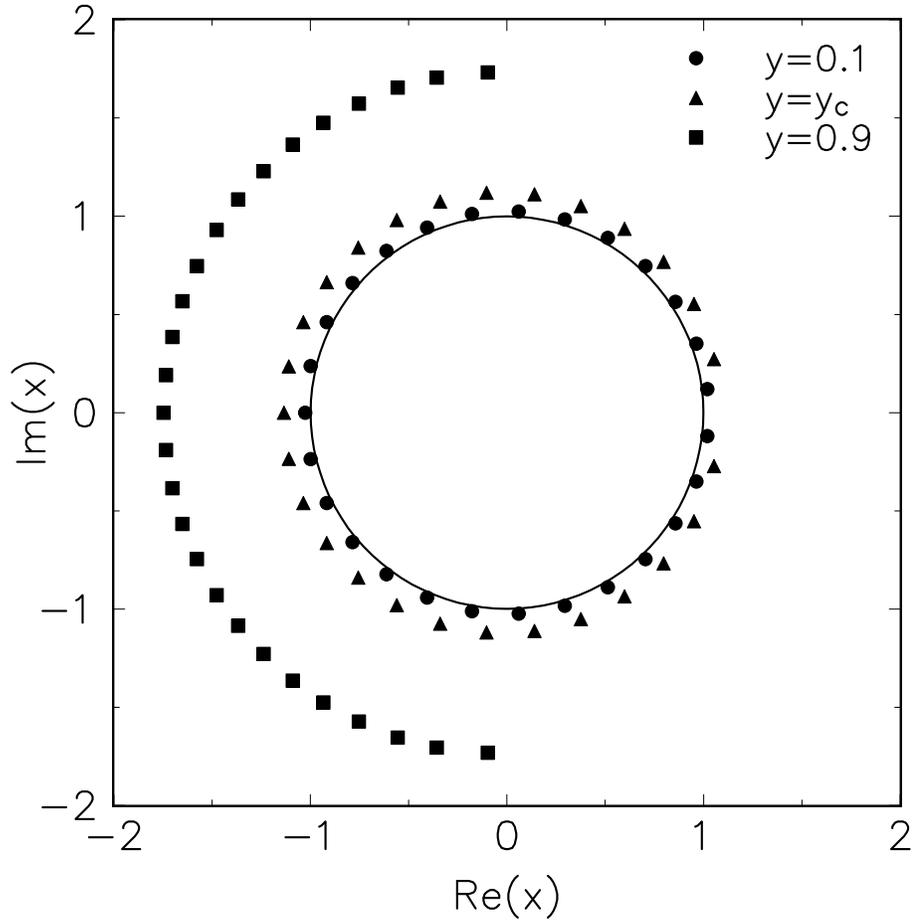}
\caption{Zeros of the three-state Potts model in the complex $x$ plane
on a $3\times3\times3$ simple-cubic lattice for $y=0.1$, 
$y=y_c\approx0.576624$ (Ref. [18]), and $y=0.9$.}
\end{figure}
 
\begin{figure}
\epsfbox{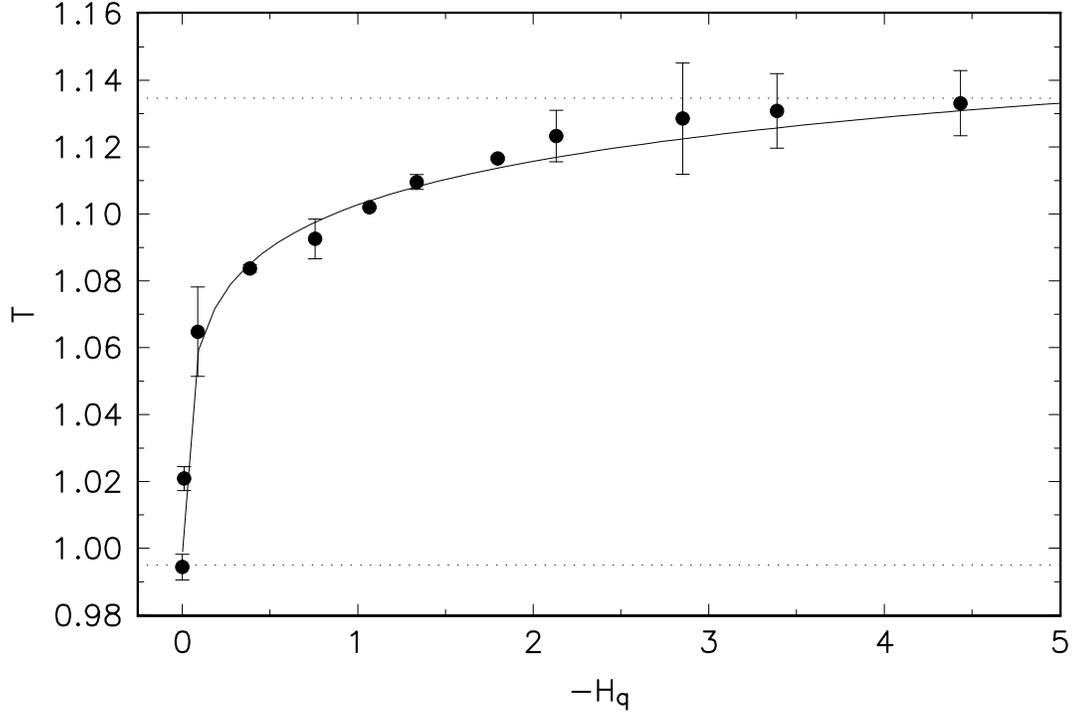}
\caption{Critical temperatures of the two-dimensional three-state Potts 
ferromagnet as a function of the magnetic field. $H_q$ is in unit of $J$
and $T$ is in unit of $J/k_B$. The upper dotted line is the Ising
transition temperature in the limit $H_q\to-\infty$, while the lower
dotted line shows the critical temperature of the three-state Potts
model for $H_q=0$.}
\end{figure} 

\end{document}